\documentclass[sigconf,nonacm]{acmart}

\usepackage[T1]{fontenc}
\usepackage[utf8]{inputenc}
\usepackage{microtype}
\usepackage{url}

\setcopyright{none}
\renewcommand\footnotetextcopyrightpermission[1]{}
\acmYear{2022}
\copyrightyear{2022}
\acmConference[Converted Manuscript]{Converted Manuscript}{2022}{}
\acmBooktitle{Converted Manuscript, 2022}

\title{Mining and Searching Association Relation of Scientific Papers Based on Deep Learning}

\author{Jie Song}
\email{songs@bupt.edu.cn}
\affiliation{%
  \institution{School of Computer Science (National Pilot School of Software Engineering), Beijing University of Posts and Telecommunications}
  \city{Beijing}
  \country{China}
}

\author{Meiyu Liang}
\authornote{Corresponding author.}
\affiliation{%
  \institution{Beijing Key Laboratory of Intelligent Telecommunication Software and Multimedia, Beijing University of Posts and Telecommunications}
  \city{Beijing}
  \country{China}
}

\author{Zhe Xue}
\affiliation{%
  \institution{Beijing Key Laboratory of Intelligent Telecommunication Software and Multimedia, Beijing University of Posts and Telecommunications}
  \city{Beijing}
  \country{China}
}

\author{Feifei Kou}
\affiliation{%
  \institution{Beijing Key Laboratory of Intelligent Telecommunication Software and Multimedia, Beijing University of Posts and Telecommunications}
  \city{Beijing}
  \country{China}
}

\author{Ang Li}
\affiliation{%
  \institution{Beijing Key Laboratory of Intelligent Telecommunication Software and Multimedia, Beijing University of Posts and Telecommunications}
  \city{Beijing}
  \country{China}
}

\begin{abstract}
There is a complex correlation among the data of scientific papers. The phenomenon reveals the data characteristics, laws, and correlations contained in the data of scientific and technological papers in specific fields, which can realize the analysis of scientific and technological big data and help to design applications to serve scientific researchers. Therefore, the research on mining and searching the association relationship of scientific papers based on deep learning has far-reaching practical significance.
\end{abstract}


\begin{document}

\maketitle

Different from the keyword matching method of the traditional search mode, the search of scientific papers is aimed at scientific researchers, which has efficient and accurate search requirements and needs to further explore the correlation relationship between scientific papers in the search results. In addition, the study relating to scientific papers faces many challenges, such as the accuracy of category segmentation, the effectiveness of entity extraction, and other problems. The traditional keyword matching search mode cannot effectively solve these problems. Feature semantic learning is carried out on the basis of existing scientific papers, so as to discover potential semantic correlations in scientific papers and establish a public semantic representation space of scientific papers from different sources. Based on interest ranking, relevance ranking, feedback mechanism, and ranking optimization mechanism, one can realize the retrieval and sorting of scientific papers and efficient partition indexing of inaccurate multi-channel information input. Furthermore, based on the deep association model, through the in-depth mining and analysis of the searched massive scientific and technological papers, the extraction and display of specialized, personalized, and orderly associations can be realized.

\section{Basic Knowledge}

A pre-trained model is a pre-trained and saved network that was previously trained on a large dataset. The basic idea of this natural language representation is used in a series of applications ranging from Word2vec\cite{mikolov2013word2vec} to BERT\cite{devlin2018bert}. Pre-trained language models are commonly used models recently. As a method of generating word vectors to pre-train neural networks, Word2vec is the first method to use pre-training natural language. The purpose of the attention mechanism is to make the model simulate the human attention mechanism to pay attention to important information. ELMo\cite{peters2018elmo} generates dynamic word vector representations according to the context on this basis, and Transformer is a feature extractor that uses the self-attention mechanism. Retrieval-oriented masked-autoencoder pre-training, exemplified by RetroMAE, further specializes language representations for matching and retrieval\cite{xiao2022retromae}. Therefore, the performance of the language model has been greatly improved. BERT pre-training uses a large amount of corpus, the general semantic representation ability is better, and the Transformer structure feature extraction ability of BERT is stronger. Chinese BERT is based on word-granularity pre-training, which can reduce the impact of unregistered words (OOV). Using position vectors to model text position information can solve the structural limitations of semantics.

Recent research shows that pre-trained language models play a crucial role in deep learning downstream tasks. Since the release of the super-large-scale pre-trained language model GPT-3\cite{dale2021gpt3} in the English domain, the training process of similar models in the Chinese domain has attracted much attention. The reach of AI beyond general language understanding is also reflected in reviews of pediatric research and clinical use\cite{li2020pediatrics}. The world's largest Chinese pre-trained language model PLUG with 27 billion parameters and 1TB+ training data has refreshed Chinese language understanding in the historical record of the evaluation benchmark CLUE classification list.

Graph representation learning comprises methods for transforming nodes, edges, and features into a lower-dimensional vector space while maximally preserving properties such as graph structure and information. Recent research shows that there are multiple ways to learn graph embedding representations, each with different levels of granularity. DeepWalk\cite{perozzi2014deepwalk} belongs to one of the graph embedding techniques using walks, a concept in graph theory that enables graph traversal by moving from one node to another as long as they are connected to common edges. Node2vec\cite{grover2016node2vec} was one of the first deep learning attempts to learn from graph-structured data. Node2vec has a walking bias variable $\alpha$ parameterized by $p$ and $q$. The parameter $p$ prioritizes the breadth-first search (BFS) process, while the parameter $q$ prioritizes the depth-first search (DFS) process. The decision of where to go next is then influenced by the probability $1/p$ or $1/q$.

As a modification of the Node2vec variant, Graph2vec\cite{narayanan2017graph2vec} essentially learns to embed subgraphs of graphs. These predetermined subgraphs have a set of edges specified by the user. Likewise, the latent subgraph embeddings are passed to the neural network\cite{li2017variance,fang2020cyclegan} for classification. Unlike previous embedding techniques, SDNE\cite{wang2016sdne} does not use random walks. Instead, it tries to learn from two different metrics: first-order proximity, where two nodes are considered similar if they share an edge, and second-order proximity, where two nodes are considered similar if they share many neighbors. LINE\cite{tang2015line} explicitly defines two functions, one for first-order approximation and the other for second-order approximation. In experiments conducted in the original study, the second-order approximation performed significantly better than the first-order, implying that including higher orders may level off the improvement in accuracy. Previous models run the risk of getting stuck in local optima because their objective functions are non-convex. HARP\cite{chen2018harp} improves the solution and avoids local optima through better weight initialization and uses graph coarsening to aggregate related nodes into ``super nodes,'' essentially a graph preprocessing step that simplifies the graph to speed up training. For scientific publications, semantic-similarity attention with hypergraph convolution supplies a higher-order representation mechanism\cite{li2026hypergraph}.

GCN\cite{kipf2016gcn} is a first-order local approximation of spectral graph convolution, which is a multi-layer graph convolutional neural network where each convolutional layer only processes first-order neighborhood information. By stacking several convolutional layers, information transfer in multi-order neighborhoods can be achieved. GCN encodes the adjacency matrix $A$ and feature matrix $X$ as embeddings $H$, which are then applied to downstream tasks. GATs\cite{velickovic2017gat} are improved on GCNs by using attention coefficients instead of Laplacian matrices. To a certain extent, GATs work better because the correlations between vertex features are better incorporated into the model. Heterogeneous graph attention further integrates typed relations with short-text semantics for semi-supervised classification\cite{hu2019han}.

Various GCN variants have been proposed to address the problem of learning graph embeddings. For example, GraphSAGE\cite{hamilton2017graphsage} samples the neighbor vertices of each vertex in the graph, aggregates the information contained in the neighbor vertices according to the aggregation function, and obtains the vector representation of each vertex in the graph for use by downstream tasks. GCN-LPA\cite{wang2020gcnlpa} analyzed the theoretical relationship between Graph Convolutional Network (GCN) and Label Propagation Algorithm (LPA), proved that edge weights that can improve the effect of LPA can also improve GCN, and then used LPA as the regularization term of GCN, which implements SOTA on the node classification task. CNMPGNN\cite{zhang2021cnmotifs} is based on the topic of common neighbors, generalizes and enriches structural patterns, groups one-hop neighbors according to CN-motifs, builds higher-order graphs, and makes full use of structural patterns. Advanced results have been achieved on several homogeneous and heterogeneous datasets. ACM-GCN\cite{luan2021acmgcn} proposes a new metric based on a similarity matrix, considering the influence of graph structure and input features on GNN, and then proposes an Adaptive Channel Mixing (ACM) framework to adaptively utilize the aggregation, diversity, and identity of channels to address harmful heterophilia. Complementary graph views can also be combined through self-supervised graph co-training for session-based recommendation\cite{xia2021graphcotraining}.

Mutual information (MI) in graphs measures the interdependence between two random variables, and much research work has been done on the mutual information of graph neural networks. DGI\cite{velickovic2019dgi} is the earliest method to apply mutual information constraints to graph-structured data, which maximizes the mutual information between the global graph summary and each of its nodes to learn informative node representations. The work of DGI has also expanded a lot. GIC\cite{mavromatis2020gic} adds a cluster-center representation\cite{sun2009knn} on the basis of DGI, modifies the loss of DGI, and significantly improves the performance of link prediction and node classification tasks. Building on DGI, DMGI\cite{park2020dmgi} devised a systematic approach to jointly integrate node embeddings from multiple graphs by introducing a consensus regularization framework that minimizes divergence between relation-type-specific node embeddings, and a universal discriminator that distinguishes real samples regardless of relation type.

However, DGI has two major limitations. First, DGI ignores the interdependencies between node embeddings and node attributes. Second, DGI does not sufficiently mine various relationships between nodes. HDMI\cite{jing2021hdmi} was proposed to address these two major limitations, designing a joint supervisory signal containing both external and internal mutual information through high-order mutual information, and optimizing the supervisory signal using high-order deep mutual information maximization. CommDGI\cite{zhang2020commdgi} proposed a Community Graph Mutual Information Maximization Network, a graph neural network designed to deal with the community detection problem, inspired by the success of deep graph mutual information maximization in self-supervised graph learning. Deep modularity-based community detection provides a complementary objective for identifying cohesive network groups\cite{yang2016community}. Similarly, GCI\cite{sun2021gci} exploits the community information of the network\cite{meng2016consensus} while using nodes as positive (or real) examples and negative (or fake) examples. When it comes to heterogeneous graphs, HDGI\cite{ren2019hdgi} is proposed and converts heterogeneous graphs into homogeneous graphs according to different meta-paths, then fuses them through an attention mechanism, and finally optimizes with DGI loss. For distributed heterogeneous graphs, reinforcement active client selection offers a way to prioritize high-contribution clients\cite{wang2025activeclient}.

There are some other graph representation learning methods that employ the theory of mutual information maximization. VIPool\cite{li2020vipool} utilizes mutual information to select the node that best represents its neighborhood, which is the only graph pooling model based on mutual information maximization. CGIPool\cite{pang2021cgipool} considers information from local neighborhoods of nodes and global dependencies between the input and the VIPool-based coarsened graph. MGNN\cite{di2020mgnn} extends the framework of GNNs by exploring aggregation and iterative schemes in mutual information methods, and proposes a new method to enlarge the normal neighborhood in graph neural network aggregation, aiming to maximize mutual information. In distributed information networks, FedSIN addresses client heterogeneity through federated self-adaptive representation learning\cite{li2026fedsin}.

Scientific paper relevance matching is to judge whether two texts are related, or to infer the relevance score between two texts. This is a central issue in the search scenario of scientific papers. Early correlation matching mainly calculates the correlation according to the text matching scores of Query and Doc. The correlation feature of word matching plays an important role in the retrieval of scientific papers. However, literal matching has its limitations. For example, it cannot handle synonyms and polysemous words. When the vocabulary is completely overlapped but the semantics expressed are completely different, literal matching cannot achieve good results. The text matching methods of deep learning mainly include representation-based matching methods and interaction-based matching methods.

Representation-based approaches use deep models\cite{li2017recursive} to represent Query and Doc respectively, and obtain semantic matching scores by calculating vector similarity. The common method is to use pre-training and migration to obtain text representations. Typical models such as the GenSen model\cite{subramanian2018gensen} can generalize sentence representations in various tasks. The Quick-thought framework\cite{logeswaran2018quickthought} uses the current sentence to predict the meaning of connected sentences and can learn sentence representations more effectively. BERT\cite{devlin2018bert} learns a good feature representation for words by running self-supervised learning methods on the basis of a massive corpus, which refreshes many natural language processing methods. Sentence-BERT\cite{reimers2019sentencebert} adopts a double or triple BERT network structure, which greatly reduces computational overhead while ensuring accuracy. To solve the problem that an unfine-tuned BERT does not perform well on text-similarity calculation, BERT-flow\cite{li2020bertflow} reversibly maps the output space of BERT from a cone to a standard Gaussian distribution space. Through normalizing flows, the BERT sentence vector distribution is reversibly mapped into a smooth, isotropic Gaussian distribution. Although the emergence of the BERT-style model has solved many discrimination problems, it is not ideal to use BERT trained directly with an unsupervised corpus for sentence representation. The Cross-Thought\cite{wang2020crossthought} model proposes an improvement to this problem. It designs a downstream task, directly optimizes the obtained sentence encoding, and uses the representation of other surrounding sentences to predict the current sentence encoding, which improves the sentence representation effect.

Another interaction-based matching method does not directly learn the semantic representation vector of Query and Doc, but allows Query and Doc to interact in advance at the bottom layer of the neural network, so as to obtain a better text vector representation, and finally passes it through a multi-layer perceptron network to obtain a semantic match score. Such models usually include an interaction layer. For example, the DIIN model\cite{gong2017diin} performs element-product interaction on the obtained Query and Doc vector encoding representations at the interaction layer, and then processes the interactive vector encoding. The MCAN model\cite{tay2018mcan} passes multiple attention calls to model multiple views to improve performance. This model introduces a multicast attention mechanism to interact and finally aggregate Query and Doc on different views. The Match$^2$ model\cite{wang2020match2} proposes a novel matching strategy, comparing the matching degree of two queries on the same Doc for similarity matching, and generates the final similarity score based on the representation and matching mode. The HCAN model\cite{rao2019hcan} also pays attention to the two matching modes to obtain the final result. It integrates the results of semantic matching, emphasizing the correspondence of meaning and the structure of components, and correlation matching, emphasizing the matching of keywords, and adopts a fully connected method to obtain the final matching result. The RE2 model\cite{yang2019re2} adopts a different design concept: keep everything simple, minimize the amount of parameters and operations, improve the inference speed, and ensure a good model effect. It designs an enhanced residual network to retain the original meaning of the text to the greatest extent and prevent it from changing in the process of network propagation\cite{xu2013imagefusion,lin2009consensus}, and uses the traditional attention mechanism to achieve interaction.

The above methods are mainly used for matching between short texts. In the search scenario of scientific papers, the length difference between Query and Doc needs to be considered. CIG-GCN-BERT\cite{liu2018cig} proposed a method to construct a long text into a concept map, interact with another long text through graph convolution, and finally get the final matching result through a multilayer perceptron\cite{li2013phd}.

\section{Semantic Feature Learning of Scientific Papers}

The feature representation of scientific papers refers to the method by which the model automatically extracts features or representations from paper data, and maps scientific papers to the same semantic vector space, thereby obtaining the semantic feature vector representation of scientific papers. The general embedding system for constructing scientific papers is mainly divided into two parts: word embedding and sentence embedding. At the scholar level, multi-view clustering with dynamic interest tracking integrates heterogeneous evidence with temporal changes in research interests\cite{li2023scholarclustering}.

Word embedding methods such as Word2vec\cite{mikolov2013word2vec} and GloVe\cite{pennington2014glove} are both unsupervised word-vector generation models. FastText\cite{yan2020bow} is an open-source, free, and lightweight library that allows users to learn text representations and text classifiers. It works on standard general-purpose hardware, and models can be downsized to fit mobile devices. To address the hyponymous relationship between words, a related word embedding projection model\cite{wang2019projection} is proposed. For visual scientific content, bi-projection fusion for omnidirectional image super-resolution provides a related example of feature enhancement\cite{wang2024omnidirectional}.

ELMo's deep contextualized word representations\cite{peters2018elmo} are learned functions of the internal states of deep bidirectional language models that model complex features of word usage, such as syntax and semantics, and how these usages vary across linguistic contexts, that is, modeling polysemy. Arora et al. propose an optimization algorithm\cite{arora2017sif} that represents sentences by a weighted average of word vectors, and provide a theoretical explanation for the success of unsupervised methods for generating sentences for models. Quick-thoughts\cite{logeswaran2018quickthought} is a simple yet effective framework for learning sentence representations from unlabeled data. As Google proposed the BERT\cite{devlin2018bert} model, pre-training models to solve the sentence embedding problem have gradually become mainstream.

However, the sentence vectors obtained by the pre-training model for similar sentences will be very different in terms of semantics. In order to solve this problem, the Sentence-BERT\cite{reimers2019sentencebert} model modifies the pre-trained BERT network, which uses Siamese and ternary network structures to derive semantically meaningful sentence embeddings that can be compared using cosine similarity. In response to the problem of inconsistent sentence embeddings expressed in multiple languages, Google proposed the multilingual BERT embedding model LaBSE\cite{annamoradnejad2020colbert}, which combines the mask language model and the translation language model and uses a bidirectional dual encoder to predict the translation ranking task, trained to generate language-independent sentence embeddings for 109 languages.

The Chinese Information Processing Institute of Beijing Normal University in China proposed a Chinese-based analogical reasoning task\cite{li2018analogical}, which provided a large CA8 Chinese corpus and a pre-trained embedding model. On the basis of early-stage Chinese language-vector research, work further explored the consistency of internal and external evaluation of word vectors and proposed a method to analyze the relationship between internal evaluation and external evaluation of Chinese character embedding\cite{qiu2018evaluations}. Compared with traditional embedding models, deep semantic representation learning\cite{keller2020semanticcode} combines entity-rich side information, such as multimodal information, knowledge graphs, and meta information, with deep models\cite{hu2018anomaly,zhao2017sliding}, such as Transformer\cite{kitaev2020reformer} and graph convolutional networks\cite{he2020lightgcn}, for deep fusion\cite{xu2013imagefusion}. Good generalization and semantic expression ability provide rich semantic capabilities for downstream application models\cite{li2014lpv}. When these representations support human decisions, interpretable machine-learning models are also important\cite{li2019interpretable}.

\section{Mining of Association Relationships of Scientific and Technological Papers}

Traditional association relation mining tasks are all based on the semantic association learning method\cite{zhang2020answer,li2017kalman}, which can discover similar relations between items in a dataset, usually focusing on the following three elements: similarity matching signal, semantic structure, and global matching. In recent years, with the development of deep learning, deep correlation models have also shown better performance than traditional methods. The DRMM\cite{guo2016drmm} model first builds local interaction features based on the basic representation vector of the text, and then uses a deep model to learn a hierarchical interaction model for matching the importance and variety of these elements.

The Fudan University team proposed the RMRN\cite{fu2020rmrn} model to solve the similarity problem of the questions recommended in community answers, and find the answerer for the question. The model was inspired by the MAC (Memory, Attention, Control) gating mechanism, and designed a reasoning memory cell, RMC (Reasoning Memory Cells), to model the question text and then conduct multi-faceted reasoning with the historical answers of the candidate users. This can mine the deep connection between the question and the user, and provide a new idea for mining the association relationship of scientific papers. Filter-enhanced MLPs provide another lightweight perspective on sequential dependencies in recommendation\cite{zhou2022filtermlp}.

Scientific papers are different from network and industry big data\cite{qian2019service} in the general sense, so the construction of a knowledge representation network model for scientific papers\cite{wang2019kg} can effectively discover the correlations in massive scientific papers. Knowledge representation of scientific papers aims to describe the entities, concepts, and their relationships existing in scientific research activities, and its essence is a complex network that reveals entity relationships. Knowledge-enhanced entity and relationship understanding has also been used to capture fine-grained semantic dependencies in sarcasm detection\cite{wang2025sarcasm}.

The problems involved in the construction of the knowledge representation network of scientific papers\cite{zhou2020reviewkg} include entity extraction of scientific papers, entity disambiguation of scientific papers, relation extraction of scientific papers, and relation inference of scientific papers. When node features and graph structure are incomplete, teacher--student distillation provides a complementary recovery strategy\cite{huo2023t2gnn}. Aiming at the missing relationship between entity nodes in the knowledge representation network, Xia Wei et al.\cite{xia2018mutualkg} used the mutual information between user-generated data to calculate the association relationship between entity nodes, and then constructed an entity association graph (EAG). According to the structure of EAG, a superposition method is proposed to calculate the potential relationship between non-adjacent entity nodes, so as to complete the knowledge graph\cite{chen2018intelligence}. Lei Jie et al.\cite{lei2020archives} established a computer-understandable scientific research archives knowledge graph semantic model, which realized the intelligent collection of archives resources, semantic organization and statistical analysis of archives big data, and explored a new path for intelligent management of scientific research archives.

With the development of projects such as public association datasets, research on association graphs\cite{yin2018relational} has gradually deepened. In the association graph, the relationship is used to connect two entities to describe the relationship between the entities, so as to form a complex association graph. Neil Veira et al.\cite{veira2019textkg} integrate textual information by adding entity embeddings and associated word embeddings to incorporate textual data into knowledge graph embeddings. This unsupervised method does not modify the optimization objective of knowledge graph embeddings, which allows it to be compatible with existing embedding models. In distributed settings, RFCSC combines dynamic client selection with adaptive gradient compression for communication-efficient reinforcement federated learning\cite{pan2025rfcsc}.

Existing methods generally use the graph network structure to mine associations. Traditional deep networks such as convolutional neural networks\cite{zhou2020universality} and recurrent neural networks\cite{sherstinsky2020rnn} are no longer applicable, but graph convolutional networks\cite{ying2018pinsage} can capture rich semantic relationships in structured data and mine the features and associations between nodes in the graph. Federated graph neural networks extend node classification across separate graphs without pooling raw graph data\cite{guan2021federatedgnn}. R-GCN\cite{schlichtkrull2018rgcn} is related to a recent class of neural networks operating on graphs and was specially developed to handle the highly multi-relational data features of real-world knowledge bases. The relational vectorized graph convolutional network VR-GCN\cite{ye2019vrgcn} simultaneously learns the embeddings of graph entities and relations for multi-relational networks. The role distinction and translation features of knowledge graphs are used in the convolution process. Since then, a VR-GCN-based alignment framework has been developed for the multi-relational network alignment task. In response to the missing problem in existing knowledge graphs, a popular method is to generate low-dimensional embedding vectors of entities and relationships and use these for inference. Shikhar Vashishth et al.\cite{vashishth2020interacte} proposed the InteractE method to enhance feature correlations between vectors in knowledge graphs and achieved the best results.

\section{Search Strategies for Scientific Papers}

Text matching methods include algorithms such as BoW\cite{yan2020bow}, TF-IDF\cite{borkakoty2020tfidf}, BM25, Jaccard\cite{bag2019jaccard}, and SimHash. Although this can mainly solve the matching problem at the lexical level, it has a great impact on matching algorithms based on lexical matching\cite{yang2015ontology}. From the application of deep learning in matching models, the models can be divided into: single-semantic models\cite{zeng2020dsdnet}, which directly calculate the similarity distance between two sentences and may lose local feature information; multi-semantic models\cite{huang2020federatedmv1}, which consider local feature information to calculate similarity from multiple perspectives\cite{xue2019multiview}; and matching-matrix models\cite{chen2020visualembedding}, in which two sentences interact first, and then the similarity is calculated through a neural network.

DSSM (Deep Structured Semantic Models)\cite{huang2013dssm} are latent semantic models with deep structure that project queries and documents into a common low-dimensional space, where the relevance of documents for a given query is easily computed as the distance between them. The proposed deep structured semantic model is discriminatively trained by maximizing the conditional likelihood of the clicked document given the query using click data. CDSSM (convolutional latent semantic model)\cite{gowthami2020cdssm} is a new latent semantic model based on a convolutional neural network for learning low-dimensional semantic vectors for search queries and web documents, which can solve the problem of the DSSM model losing contextual information. H. Palangi et al.\cite{palangi2014lstm} proposed the use of LSTM-DSSM (Long-Short-Term Memory) to capture long-distance context features, which is a deep structured semantic model or deep semantic similarity model (DSSM) and LSTM network. A Semantics-adversarial and Media-adversarial Cross-media Retrieval method (SMCR)\cite{li2022smcr} is proposed to minimize the loss of intra-media discrimination, inter-media consistency, and intra-semantics discrimination.

For the multi-view problem of matching sentences, the MV-DSSM\cite{huang2020federatedmv1} model was proposed to solve this problem from the information source, while the Chinese Academy of Sciences team proposed a new deep architecture\cite{wan2015positional} to match two sentences with multiple positional sentence representations. Supervised cross-modal retrieval under federated learning extends this setting to decentralized heterogeneous data\cite{li2024federatedretrieval}. Specifically, each location sentence representation is the sentence representation of that location, generated by bidirectional long short-term memory (Bi-LSTM), which exploits the rich context of the entire sentence to capture the contextual local information in each location sentence representation. The interactions between these different positional sentence representations are aggregated through $k$-Max pooling and multilayer perceptrons, resulting in a matching score.

Relevance can be abstracted to a certain extent as the semantic similarity between Doc and Query. Current research on semantic similarity\cite{kou2016social} is very mature. Jun Xu et al.\cite{xu2018matching} conducted a comprehensive review of deep learning matching in search and recommendation. In terms of semantic matching, the focus is often on how to define ``matching,'' especially how to better match the encoding content of the two. However, in the application process, there is still a big gap between relevance and similarity, because small changes in search content will bring large changes in semantic intent. Recent generative recommendation work addresses stage-wise information loss by unifying retrieval and ranking in a single model\cite{zhang2025unifiedretrieval}.

Thanh V. Nguyen et al. proposed the QUARTS model\cite{nguyen2020quarts}, a deep end-to-end model that learns to efficiently classify mismatches and generate mismatched examples to improve the classifier by using real samples and generated samples. A latent variable is introduced in the alternating cross-entropy loss to train the model end-to-end. This not only makes the classifier more robust, but also improves the overall ranking performance. As the amount of data grows, some pre-trained semantic models introduce two pre-training tasks: MLM (Masked Language Model)\cite{salazar2019mlm} and NSP (Next Sentence Prediction)\cite{liu2019nsp}. Examples include Baidu ERNIE\cite{sun2019ernie}, Tsinghua ERNIE\cite{zhang2019ernie}, and K-BERT\cite{liu2020kbert}. These models incorporate more external knowledge.

\section{Conclusion}

The search and correlation mining of scientific papers is an unavoidable process for researchers to carry out scientific research work. The search of scientific papers is different from the keyword matching method of the traditional search mode. It has efficient and accurate search requirements, and it is necessary to further explore the correlation in the scientific papers in the search results. Therefore, it is necessary to carry out feature semantic learning based on the existing scientific papers, to discover potential semantic associations in scientific papers, thereby recalling higher-quality search results.

\section*{Acknowledgement}


\end{document}